\documentclass[12pt,a4paper]{article}
\pdfoutput=1
\usepackage{amsmath,amssymb,exscale}
\usepackage{amsfonts,amsthm}
\usepackage{graphicx, float, epsfig}
\usepackage{mathrsfs}
\usepackage{slashed}
\usepackage{textcomp}
\usepackage[T1]{fontenc}
\usepackage[utf8]{inputenc}
\usepackage[table]{xcolor}
\usepackage[lowtilde]{url}
\usepackage{cancel}
\numberwithin{equation}{section}
\usepackage{verbatim} 
\usepackage{appendix}
\usepackage{hyperref}

\newcommand{\be}{\begin{equation}}
\newcommand{\ee}{\end{equation}}
\newcommand{\ba}{\begin{eqnarray}}
\newcommand{\ea}{\end{eqnarray}}
\newcommand{\bay}{\begin{array}{rcl}}
\newcommand{\eay}{\end{array}}
\newcommand{\ra}{\rightarrow}

\newcommand{\bra}{\langle}
\newcommand{\ket}{\rangle}

\def\LP{l_{\rm Pl}}

\def\MP{M_{\rm Pl}}

\begin{document}

\title{A Chern-Simons model for baryon asymmetry}

\author{Risto Raitio\footnote{E-mail: risto.raitio@gmail.com}\\	
02230 Espoo, Finland}

\date{April 3, 2023}  \maketitle 

\abstract{\noindent
In search of a phenomenological model that would describe physics from Big Bang to the Standard Model (SM), we propose a model with the following properties (i) above an energy about $\Lambda_{cr} > 10^{16}$ GeV there are Wess-Zumino supersymmetric preons and Chern-Simons (CS) fields, (ii) at $\Lambda_{cr} \sim 10^{16}$ GeV spontaneous gauge symmetry breaking takes place in the CS sector and the generated  topological mass provides an attractive interaction to equal charge preons, (iii) well below $10^{16}$ GeV the model reduces to the standard model with essentially pointlike quarks and leptons, having a radius $~\sim 1/\Lambda_{cr} \approx 10^{-31}$ m. The baryon asymmetry turns out to have a fortuitous ratio $n_B/n_{\gamma} \ll 1$.}

\vskip 1.0cm
\noindent


\vskip 0.5cm
\noindent
DOI: https://doi.org/10.1016/j.nuclphysb.2023.116174

\newpage

\tableofcontents
\vskip 1.0cm

\section{Introduction}
\label{intro}

The observation of expanding baryon asymmetric universe is about 100 years old. The concordance (standard) model \cite{astrocosmo} has been developed explaining observations though not with the same precision and extension as in particle physics \cite{particle}. Even missing pieces exist like dark matter, baryon asymmetry and quantum gravity.

In this note we take a first step to draft a model for both particles and cosmology with simplicity as the main principle of unification. It is commonly understood that the quark electric charges and running interaction coupling constants in the standard model (SM) imply a large unified gauge group with rich spectra of particles. We take an alternative position of keeping number of elementary particles small, determined by global supersymmetry, and fermions obeying Dirac equation with SM gauge interactions. In addition, we reinforce our previous  model by topological concepts of Chern-Simons model. On the other hand, we do not exclude any current structure, like string theory, loop quantum gravity, etc. but we want to start from certain, in our opinion, simpler concepts and see how far they can take us. 

So we split quarks and leptons in three pointlike constituents, called in this note chernons (synonym for preon\footnote{The term was coined by Pati and Salam in 1974 \cite{Pati_S}.} or superon). The reason for doing so is disclosed in section \ref{barasym}. Of the many preon models in the literature there are two of them which are like ours. One of them was proposed by Harari, and independently by Shupe \cite{Harari, Shupe}. The model of Finkelstein \cite{Finkelstein} was developed from a different basis, including the quantum symmetry group SLq(2) and knot theory. It turned out, however, to agree with \cite{Harari, Shupe}. The major difference between the above models and our model \cite{Rai_00, Rai_01} is that ours has its basis in unbroken global supersymmetry where superpartners are in the model initially, not as new sparticles to be found in the future. 

The scale where three chernon bound states form, making the standard model particles in 1+2 dimensions, is assumed to be near the usual grand unified theory (GUT) scale, about $10^{16}$ GeV, denoted here by $\Lambda_{cr}$. Below $\Lambda_{cr}$ all the four preon scenarios of the previous paragraph revert to the standard model at accelerator energies. Above  $\Lambda_{cr}$ in the early universe chernons were momentarily located nearly fixed in the comoving frame of the rapidly expanding universe making the 1+2 dimensional potential energy description a reasonable approximation.

Chern-Simons-Maxwell (CSM) models (\ref{mcsaction}) have been studied in condensed matter physics papers, e.g. \cite{Deser_J_T, Giro_G_d_M_N, Beli_D_F_H}. In this note we apply the CSM model in particle physics phenomenology at high energy in the early universe. 

We construct the visible matter of two fermionic chernons: (i) one charged $m^-$, (ii) one neutral $m^0_V$, V = R, G, B, carrying quantum chromodynamics (QCD) color, and the photon $A$. The action is C symmetric. The chernons have zero (or small) mass. Weak interactions operate below $\Lambda_{cr}$ between quarks and leptons, just as in SM. The chernon baryon (B) and lepton (L)  numbers are zero. Given these quantum numbers, quarks consist of three chernons, as indicated in table \ref{tab:table0}. There could be more composite states like those containing $m^+m^-$ pair. This annihilates immediately into other particles, which form later leptons and quarks.

\begin{table}[h!]
	\begin{center}
		\begin{tabular}{|l|l|} 
			\hline
			SM quark & chernon state \\ 
			\hline 
			$u_R$ & $m^+ m^+ m^0_R$ \\				 
			$u_G$ & $m^+ m^+ m^0_G$ \\  			
			$u_B$ & $m^+ m^+ m^0_B$ \\
			\hline
			$d_R$ & $m^- m^0_G m^0_B $ \\		
			$d_G$ & $m^- m^0_B m^0_R$ \\			
			$d_B$ & $m^- m^0_R m^0_G$ \\
			\hline
		\end{tabular}
		\caption{\footnotesize{Quark-chernon correspondence. The upper index of $m$ is charge zero or $\pm \frac{1}{3}$. The lower index is color $R, G$ or $B$.}}
		\label{tab:table0}
	\end{center}
\end{table}

The article is organized as follows. In section \ref{kinetic} we recap the Wess-Zumino model kinetic terms of the supersymmetric chernons and some scalars. The full Chern-Simons-QED$_3$ action is given in section \ref{mcsqed3action}. The chernon-chernon interaction potential is disclosed in \ref{potential}. The transformation from chernons to quarks and leptons takes place during inflation, which is briefly reviewed in section \ref{inflation}. Sakharov conditions are discussed in section{sakharov}. Our mechanism for baryon asymmetry is proposed in section \ref{barasym}. Conclusions are given in section \ref{conclusions}.  In the appendix \ref{appdx} a table of visible and dark matter is displayed.

\section{Wess-Zumino action kinetic terms}
\label{kinetic}

We briefly recap our chernon (superon) scenario of \cite{Rai_00, Rai_01}, which turned out to have close resemblance to the simplest $N=1$ globally supersymmetric 1+3 model, namely the free, massless Wess-Zumino model \cite{Wess_Z, Figueroa-O} with the kinetic Lagrangian including three neutral fields $m$, $s$, and $p$ with $J^P = \frac{1}{2}^+, 0^+$, and $0^-$, respectively
\be
\mathcal{L}_{\rm{WZ}} = -\frac{1}{2} \bar{m}\cancel\partial m - \frac{1}{2} (\partial s)^2 - \frac{1}{2} (\partial p)^2
\label{chirallagrangian}
\ee
where $m$ is a Majorana spinor, $s$ and $p$ are real fields (metric is mostly plus). 

We assume that the pseudoscalar $p$ is the axion \cite{Peccei_Q}, and denote it below as $a$. It has a fermionic superparther, the axino $n$, and a bosonic superpartner, the saxion $s^0$.

In order to have visible matter we assume the following charged chiral field Lagrangian
\be
\mathcal{L}_{-} = -\frac{1}{2}m^- \cancel\partial m^- -\frac{1}{2}(\partial s^-_i)^2, ~~i=1,2
\label{chargedchiral}
\ee

\section{Chern-Simons-QED$_3$ action}
\label{mcsqed3action}

A number of 1+2 dimensional models have properties close to 1+3 dimensional world as can be found in \cite{Deser_J_T, Jackiw, Malda}, see also \cite{Liang_Z}. Our choice here is 1+2 dimensional Chern-Simons (CS) action is \cite{cs, Witten_0}
\be
S = \frac{k}{4\pi}\int_M \rm{tr}(A\wedge dA + \frac{2}{3}A\wedge A\wedge A)
\label{mcsaction}
\ee
where $k$ is the level of the theory and $A$ the connection. (The 	compatibility of different dimensions is discussed in section \ref{inflation}.)

The action for a Chern-Simons-QED$_{3}$ model \cite{Beli_D_F_H, N.Cimento} including two polarization $\pm$ fermionic fields ($\psi _{+},\psi _{-}$), a gauge field $A_{\mu }$ and a complex scalar field $\varphi$ with spontaneous breaking of local U(1) symmetry is

\begin{align}
	S_{\rm{CS-QED_3}} = \int d^{3}x\{-\frac{1}{4}F^{\mu \nu }F_{\mu \nu }+i\overline{%
		\psi }_{+}\gamma ^{\mu }D_{\mu }\psi _{+}+i\overline{\psi }_{-}\gamma ^{\mu
	}D_{\mu }\psi _{-} \nonumber \\ 
	+ {\frac12}\theta \epsilon ^{\mu v\alpha }A_{\mu }\partial _{v}A_{\alpha }-m_{e}(%
	\overline{\psi }_{+}\psi _{+}-\overline{\psi }_{-}\psi _{-})  \nonumber \\
	- y(\overline{\psi }_{+}\psi _{+}-\overline{\psi }_{-}\psi _{-})\varphi
	^{\ast }\varphi +D^{\mu }\varphi ^{\ast }D_{\mu }\varphi -V(\varphi ^{\ast
	}\varphi )\},  
	\label{actionMCS}
\end{align}
where the covariant derivatives are $D_{\mu }\psi _{\pm }=(\partial _{\mu }+ie_{3}A_{\mu })\psi_{\pm }$ \ and $D_{\mu }\varphi =(\partial _{\mu }+ie_{3}A_{\mu })\varphi$. $\theta$ is the important topological parameter and $e_{3}$ is the coupling constant of the $U(1)$ local gauge symmetry, here with dimension of $(\rm{mass})^{1/2}$.
$V(\varphi ^{\ast }\varphi )$ represents the self-interaction potential,
\begin{equation}
	V(\varphi ^{\ast }\varphi )=\mu ^{2}\varphi ^{\ast }\varphi +\frac{\zeta }{2}%
	(\varphi ^{\ast }\varphi )^{2}+\frac{\lambda }{3}(\varphi ^{\ast }\varphi
	)^{3}
	\label{potential6}
\end{equation}
which is the most general  sixth power renormalizable potential in 1+2 dimensions \cite{Delcima}. The parameters $\mu,~ \zeta,~ \lambda $ and $y$ have mass dimensions 1, 1, 0 and 0, respectively. For potential parameters $\lambda >0, \zeta <0$ and $\mu ^{2}\leq 3\zeta ^{2}/(16\lambda)$  the vacua are stable.  

In 1+2 dimensions, a fermionic field has its spin polarization fixed up by the sign of mass  \cite{Binegar}. The model includes two positive-energy spinors (two spinor families). Both of them obey Dirac equation, each one with one polarization state according to the sign of the mass parameter. 

The vacuum expectation value $v$ of the scalar field $\varphi$ is given by: 
\be
\langle \varphi ^{\ast }\varphi \rangle =v^{2}=-\zeta /\left( 2\lambda
\right) +\left[ \left( \zeta /\left( 2\lambda \right) \right) ^{2}-\mu
^{2}/\lambda \right] ^{1/2}
\label{vev}
\ee
The condition for its minimum is $\mu ^{2}+\frac{\zeta }{2}%
v^{2}+\lambda v^{4}=0$. \ After the spontaneous symmetry breaking, the
scalar complex field can be parametrized by $\varphi =v+H+i\theta $, where $
H$ represents the Higgs scalar field and $\theta $ the would-be Goldstone
boson. For manifest renormalizability one adopts the 't Hooft gauge by adding the gauge fixing term $S_{R_{\xi}}^{gt}=\int d^{3}x[-\frac{1}{2\xi }(\partial ^{\mu }A_{\mu }-\sqrt{2}\xi M_{A}\theta)^{2}]$ to the broken action. Keeping only the bilinear and the Yukawa interaction terms one has the following action

\begin{align}
	{S}_{{\rm {CS-QED}}}^{{\rm SSB}} & =\int d^{3}x\biggl\{-\frac{1}{4}F^{\mu \nu}
	F_{\mu \nu }+\frac{1}{2}M_{A}^{2}A^{\mu }A_{\mu } \nonumber \\
	& -\frac{1}{2\xi }(\partial^{\mu }A_{\mu })^{2}+\overline{\psi }_+(i\cancel\partial -m_{eff})\psi _{+} \nonumber \\
	& +\overline{\psi }_{-}(i\cancel\partial +m_{eff})\psi _{-}+ \frac{1}{2}
	\theta \epsilon ^{\mu v\alpha }A_{\mu }\partial _{v}A_{\alpha }  \nonumber \\
	& +\partial ^{\mu }H\partial _{\mu }H-M_{H}^{2}H^{2} +\partial ^{\mu }\theta
	\partial _{\mu }\theta -M_{\theta }^{2}\theta ^{2} \nonumber \\
	& -2yv(\overline{\psi }_{+}\psi _{+}-\overline{\psi }_{-}\psi _{-})H-e_{3}\left( \overline{\psi }
	_{+}\cancel A\psi _{+}+\overline{\psi }_{-}\cancel A\psi _{-}\right) \biggr\}  \label{actionMCS3}
\end{align}
where the mass parameters 
\begin{equation}
	M_{A}^{2}=2v^{2}e_{3}^{2},~~m_{eff}=m_{e}+yv^{2},~~M_H^{2}=2v^{2}(\zeta +2\lambda v^{2}),~~M_{\theta }^{2}=\xi M_{A}^{2}
	\label{}
\end{equation}
depend on the SSB mechanism. The Proca mass, $M_{A}^{2}$originates from the Higgs mechanism. The Higgs mass, $M_{H}^{2}$, is associated with the real scalar field. The Higgs mechanism also contributes to the chernon mass, resulting in an effective mass \ $m_{eff}$. There are two photon mass-terms in (\ref{actionMCS3}), the Proca and the topological one. 

\section{Chernon-Chernon interaction}
\label{potential}

The chernon-chernon scattering amplitude in the non-relativistic approximation is obtained by calculating the t-channel exchange diagrams of the Higgs scalar and the massive gauge field. The propagators of the two exchanged particles and the vertex factors are calculated from the action (\ref{actionMCS3}) \cite{Beli_D_F_H}.

The gauge invariant effective potential for the scattering considered is obtained in \cite{Kogan, Dobroliubov}
\begin{equation}
	V_{{\rm CS}}(r)=\frac{e^{2}}{2\pi }\left[ 1-\frac{\theta }{m_{e}}\right]
	K_{0}(\theta r)+\frac{1}{m_{e}r^{2}}\left\{ l-\frac{e^{2}}{2\pi \theta }%
	[1-\theta rK_{1}(\theta r)]\right\} ^{2} 
\label{Vmcs}
\end{equation}
where $K_{0}(x)$ and $K_{1}(x)$ are the modified Bessel functions and $l$ is the angular momentum ($l=0$ in this note). In (\ref{Vmcs}) the first term $[~]$ corresponds to the electromagnetic potential, the second one $\{~\}^2$ contains the centrifugal barrier $\left(l/mr^{2}\right)$, the Aharonov-Bohm term and the two photon exchange term.

One sees from (\ref{Vmcs}) the first term may be positive or negative while the second term is always positive. The function $K_{0}(x)$ diverges as $x \ra 0$ and approaches zero for $x \ra \infty$ and $K_{1}(x)$ has qualitatively similar behavior. For our scenario we need negative potential between equal charge chernons. Being embarrassed of having no data points for several parameters in (\ref{Vmcs}) we can give one relation between these parameter values for a negative potential. We must have the condition\footnote{For applications to condensed matter physics, one must require $\theta \ll m_{e}$, and the scattering potential given by (\ref{Vmcs}) then comes out positive  \cite{Beli_D_F_H}.}
\be
\theta \gg m_e
\label{condition}
\ee
The potential (\ref{Vmcs}) also depends on $v^{2}$, the vacuum expectation value, and on $y$, the parameter that measures the coupling between fermions and Higgs scalar. Being a free parameter, $v^{2}$ indicates the energy scale of the spontaneous breakdown of the $U(1)$ local symmetry.

\section{Inflation and Supergravity}
\label{inflation}


We discuss briefly, and in simple terms, the question of different dimensions of CS theory and gravity. We assume that the universe at $t \sim 0$ included a subspace of one dimension less than the manifold of general relativity $M_{GR}$.\footnote{A line is one dimensional when looked from a distance but by getting very close to it one sees, or rather knows, it consists of zero dimensional points, that is numbers.} A promising example of such a theory is Chern-Simons gauge theory defined in a smooth, compact three-manifold $M_{CS} \subset M_{GR}$, having a gauge group $G$, which is semi-simple and compact, and an integer parameter $k$. The Chern-Simons field equations (\ref{mcsaction}) require that $A$ be flat \cite{Witten_0}. The curvature tensor may be decomposed, in any spacetime dimension, into a curvature scalar $R$, a Ricci tensor $R_{\mu\nu}$, and a conformally invariant Weyl tensor $C_{\mu\nu\rho}^{~~~~\sigma}$. In 1+2 dimensions the Weyl tensor vanishes identically, and the Riemann curvature tensor $R_{\mu\nu\rho\sigma}$ is determined algebraically by the curvature scalar and the Ricci tensor. Therefore any solution of the vacuum Einstein field equations is flat and any solution of the field equations with a cosmological constant $R_{\mu\nu} = 2\Lambda g_{\mu\nu}$ has constant curvature. Physically, a 1+2 dimensional spacetime has no local degrees of freedom. There are no gravitational waves in the classical theory, and no gravitons in the quantum theory

CS theory, defined earlier by the action (\ref{mcsaction}), is a topological, quantizable gauge field theory \cite{Witten_0}. The appropriate observables lead to vevs which correspond to topological invariants. The observables have to be gauge invariant. Secondly, they must be independent of the metric. Wilson loops verify these two properties \cite{Witten_0}, and they are therefore the key to observables to be considered in Chern-Simons theory. Independence of metric gives CS theories the desireable property of background independence. The CS interaction (\ref{mcsaction}) is effective only at energy scales near and above $\Lambda_{cr}$. This we interpret as chernons living (mod 3) on surfaces of spheres with diameter of the order of $1/\Lambda_{cr}$. These composite states are quarks and leptons of the standard model in 1+3 dimensions.

In summary, the potential (\ref{Vmcs}) dominates over general relativity, and Coulomb repulsion, at distances below $1/\Lambda_{cr}$ in the 1+2 dimensional manifold $M_{CS}$ while at larger distances gravity is stronger. 

At the beginning of inflation, $t = t_i \sim 10^{-36}$ s, the universe is modeled by 1+3 dimensional classical gravity, and Chern-Simons theory as long as $T \geq \Lambda_{cr}$. The Einstein-Hilbert action is 
\be
S = \int d^4 x \sqrt{-g}\Big(\frac{1}{2}R - \frac{1}{2}g^{\mu\nu}\partial_{\mu}\phi \partial_{\nu}\phi - V(\phi)\Big)
\label{einsteinhilbertscalar}
\ee
\noindent
The E-H action dominates rapidly leading inflation to end at $t_R \approx 10^{-32}$ s. Then the inflaton, which is actually coherently oscillating homogeneous field, a Bose condensate, reaches the minimum of its potential. There it oscillates and decays to SM particles produced from chernons in the earlier phase of inflation. This causes the reheating phase, or the Bang, giving visible matter particles more kinetic energy than dark matter particles have. 

The CMB measurements of inflation can be well described by a few simple slow-roll single scalar potentials in (\ref{einsteinhilbertscalar}). One of the best fits to Planck data \cite{Planck2018} is obtained by one of the very oldest models, the Starobinsky model \cite{Starobinsky}. The action is
\be
S = \frac{1}{2} \int d^4 x \sqrt{-g}\Big(R + \frac{R^2}{6M^2}\Big)
\label{einsteinhilbertrscalarsquared}
\ee
where $M \ll \MP$ is a mass scale. Current CMB measurements indicate scale invariant spectrum with a small tilt in scalar density $n_s = 0.965 \pm 0.004$ and an upper limit for tensor-to-scalar ratio $r < 0.06$. These values are fully consistent with the Starobinsky model (\ref{einsteinhilbertrscalarsquared}) which predicts $r \simeq 0.003$. 

The model (\ref{einsteinhilbertrscalarsquared}) has the virtue of being based on gravity only physics. Furthermore, the Starobinsky model has been shown to correspond to no-scale supergravity coupled to two chiral supermultiplets. Some obstacles have to be sorted out first before reaching supergravity. To do that we follow the review by Ellis et al. \cite{Ellis_G_N_N_O_V}.

The first problem with generic supergravity models with matter fields is that their effective potentials do not provide slow-roll inflation as needed. Secondly, they may have anti-deSitter vacua instead of deSitter ones. Thirdly, looking into the future, any new model of particles and inflation should preferably be consistent with some string model properties. These problems can be overcome by no-scale supergravity models. No-scale property comes from their effective potentials having flat directions without specific dynamical scale at the tree level. This has been derived from string models, whose low energy effective theory supergravity is.

Other authors have studied other implications of superstring theory to inflationary model building focusing on scalar fields in curved spacetime \cite{Guth_M_K} and the swampland criteria \cite{Vafa_B_C, Brandenberger_H_B_R, Brandenberger_J_H}. These studies point out the inadequacy of slow roll single field inflation. We find it important to establish first a connection between the Starobinsky model and (two field) supergravity.

The bosonic supergravity Lagrangian includes a Hermitian function of complex chiral scalar fields $\phi_i$ which is called the K\"{a}hler potential $K(\phi^i,\phi^*_j)$. It describes the geometry of the model. In minimal supergravity (mSUGRA) $K = \phi^i \phi^*_i$. Secondly the Lagrangian includes a holomorphic function called the superpotential $W(\phi^i)$. This gives the interactions among the fields $\phi^i$ and their fermionic partners. $K$ and $W$ can be combined into a function $G \equiv K + \ln|W|^2$. The bosonic Lagrangian is of the form
\be
\mathcal{L} = -\frac{1}{2}R + K^j_i \partial_{\mu}\phi^i\partial^{\mu}\phi^*_j - V - \frac{1}{4}\rm{Re}(f_{\alpha\beta})F^{\alpha}_{\mu\nu}F^{\beta\mu\nu} - \frac{1}{4}\rm{Im}(f_{\alpha\beta})F^{\alpha}_{\mu\nu}\tilde{F}^{\beta\mu\nu}
\label{bosonlagrangian}
\ee
where $K^j_i \equiv \partial^2 K/\partial\phi^i \partial\phi^*_j$ and $\rm{Im}(f_{\alpha\beta})$ is the gauge kinetic function of the chiral fields $\phi^i$. In mSUGRA the effective potential is
\be
V(\phi^i,\phi^*_j) = e^K\big[|W_i + \phi^*_iW|^2 - 3|W|^2\big]
\label{msugrapotential}
\ee
where $W_i \equiv \partial W/\partial\phi^i$. It is seen in (\ref{msugrapotential}) that the last term with negative sign may generate AdS holes with depth $-\mathcal{O}(m^2_{3/2}\MP^2)$ and cosmological instability. Solution to this and the slow-roll problem is provided by no-scale supergravity models. The simplest such model is the single field case with
\be
K = -3\ln(T + T^*)
\label{ksinglefield}
\ee
where $T$ is a volume modulus in a string compactification. 

The single field (\ref{ksinglefield}) model can be generalized to include matter fields $\phi^i$ with the followng K\"{a}hler potential
\be
K = -3\ln(T + T^* - \frac{1}{3}|\phi_i|^2)
\label{multifield}
\ee

The no-scale Starobinsky model is now obtained with some extra work from the potential (\ref{msugrapotential}) and assuming $\bra T \ket = \frac{1}{2}$. For the superpotential the Wess-Zumino form is introduced \cite{Ellis_N_O}
\be
W = \frac{1}{2}M \phi^2 - \frac{1}{3}\lambda \phi^3
\label{wzsuperpotential}
\ee
which is a function of $\phi$ only. Then $W_T = 0$ and from $V' = |W_{\phi}|^2$ the potential becomes as
\be
V(\phi) = M^2\frac{|\phi|^2 |1 -\lambda\phi/M|^2}{(1 - |\phi|^2/3)^2}
\label{wzpotential}
\ee
The kinetic terms in the scalar field Lagrangian can be written now
\be
\mathcal{L} = (\partial_{\mu}\phi^*, \partial_{\mu}T^*)\Big(\frac{3}{(T + T^* - |\phi|^2/3)^2}\Big) \left( \begin{array}{cc} (T+T^*)/3 & -\phi/3 \\
	-\phi^*/3 & 1 \end{array} \right) \left( \begin{array}{c} \partial^{\mu}\phi \\ \partial^{\mu}T \end{array} \right)
\label{}
\ee
Fixing $T$ to some value one can define the canonically normalized field $\chi$
\be
\chi \equiv \sqrt{3}\tanh^{-1}\Bigg(\frac{\phi}{\sqrt{3}}\Bigg)
\label{}
\ee
By analyzing the real and imaginary parts of $\chi$ one finds that the potential (\ref{wzpotential}) reaches its minimum for Im$\chi = 0$. Re$\chi$ is of the same form as the Starobinsky potential in conformally transformed Einstein-Hilbert action \cite{Stelle} with a potential of the form 
\be
V = \frac{3}{4}M^2(1 - e^{-\sqrt{2/3}\phi})^2 
\label{key}
\ee
when 
\be
\lambda = \frac{M}{\sqrt{3}}
\label{}
\ee
Most interestingly, $\lambda/M$ has to be very accurately $1/\sqrt{3}$, better than one part in $10^{-4}$, for the potential to agree with measurements. 

This is briefly the basic mechanism behind inflation in the Wess-Zumino mSUGRA model, which foreruns reheating of visible matter. But only the particles containing $m$ chernons, i.e. the visible matter gets reheated. The dark sector is going through reheating unaffected and is distributed smoothly all over space. The quantum fluctuations of the dark fields are enhanced by gravitation and provide a clumpy underlay for visible matter to form objects of various sizes, from stars to large scale structures.

\section{Sakharov conditions}
\label{sakharov}

Sakharov suggested \cite{Sakharov} three necessary conditions that must satisfied to produce matter and antimatter at different rates. They are (i) baryon number B  violation, (ii) C-symmetry and CP-symmetry violation and (iii) interactions out of thermal equilibrium.

Baryon number violation is clearly needed to reach baryon asymmetry. This is valid in our model because baryon number is not defined conventionally. C-symmetry violation is needed so that the interactions which produce more baryons than anti-baryons will not be counterbalanced by interactions which produce more anti-baryons than baryons. This is discussed in section \ref{nmdiff}. CP-symmetry violation is required because otherwise equal numbers of left-handed baryons and right-handed anti-baryons would be produced, as well as equal numbers of left-handed anti-baryons and right-handed baryons. The observed pattern of CP-violation \cite{astrocosmo} remarkably confirms the Cabibbo–Kobayashi–Maskawa (CKM) description of three fermionic generations of particles \cite{Cabibbo, Koba_M}. CP-violation phenomenology is discussed in detail in \cite{Bonn_G_G_R_1, Bonn_G_G_R_2}. Our present one generation "skeleton" model cannot satisfy this condition but in principle, by completing the model and deriving the low energy limit, it could be explained. In the SM, the CKM model gives an explanation of why the breaking is so small, despite the phase associated to it being of order one. Thirdly, interactions are out of thermal equilibrium in a rapidly expanding universe.

\section{Baryon asymmetry}
\label{barasym}

We now examine the potential (\ref{Vmcs}) in the early universe. Consider large number of groups of twelve chernons each group consisting of four $m^+$, four $m^-$ and four $m^0$ particles. Any bunch may form only electron and proton (hydrogen atoms H), only positron and antiproton ($\bar{\rm{H}}$) or some combination of both H and $\bar{\rm{H}}$ atoms \cite{Rai_00, Rai_01}. This is achieved by arranging the chernons appropriately (mod 3) using table \ref{tab:table1}. This way the transition from matter-antimatter symmetric universe to matter-antimatter asymmetric one happens straightforwardly.

Because the Yukawa force (\ref{Vmcs}) is the strongest force the light $e^-$, $e^+$ and the neutrinos are expected to form first at the very onset of inflation. To obey condition $B-L=0$ of baryon-lepton balance and to sustain charge conservation, for one electron made of three chernons, nine other chernons have to be created simultaneously, these form a proton. Accordingly for positrons. One neutrino requires a neutron to be created. The $m^0$ carries in addition color enhancing neutrino formation. This makes neutrinos different from other leptons and the quarks. 

Later, when the protons were formed, because chernons had the freedom to choose whether they are constituents of $\rm{H}$ or $\bar{\rm{H}}$ there are regions of space of various sizes dominated by $\rm{H}$ or $\bar{\rm{H}}$ atoms. Since the universe is the largest statistical system it is expected that there is only a very slight excesses of $\rm{H}$ atoms (or $\bar{\rm{H}}$ atoms which only means a charge sign redefinition) which remain after the equal amounts of $\rm{H}$ and $\bar{\rm{H}}$ atoms have annihilated. The ratio $n_B/n_{\gamma}$ is thus predicted to be $\ll 1$. The ratio $n_B/n_{\gamma}$ is a multiverse-like concept.

Fermionic dark matter has in this scenario no mechanism to become "baryon" asymmetric like visible matter. Therefore we expect that part of fermionic dark matter has annihilated into bosonic dark matter. Secondly, we predict there should exist both dark matter and anti-dark matter clumps attracting visible matter in the universe. Collisions of anti-dark matter and dark matter celestial bodies would give us a new source for wide spectrum gravitational wave production (the lunar mass alone is $\sim 10^{49}$ GeV).

\section{Nucleon isospin violation}
\label{nmdiff}

The topological mass works in favor of heavier d-quark and neutron \cite{Rai_03}, in qualitative agreement with lattice calculations. Care must be taken not to do double counting for d/u-quark mass difference with respect to CS-QED$_3$ calculations and QCD/QED lattice results. It is plausible that the topological terms in action (\ref{actionMCS}) are very small on scales $\ll \Lambda_{cr}$ in 1+3 dimensions and therefore QCD/QED only contribute to the mass difference.

\section{Conclusions and Outlook}
\label{conclusions}

Above $\Lambda_{cr}$ the fermionic chernons are C symmetric with equal masses and charges symmetrically around zero: \{-1/3, 0, 1/3\}. Below the transition energy $\Lambda_{cr}$ fractional charge chernon composites form quarks while charge zero and one states are leptons as shown in table \ref{tab:table1}. These composite states behave to a good approximation like pointlike particles: the composite radius being of the order of $10^{-31}$ m corresponding to a photon energy of $\Lambda_{cr} \sim 10^{16}$ GeV. Below this energy the standard model is obtained \cite{Harari, Shupe, Finkelstein, Rai_01} and photons lose their resolving power to differentiate the Yukawa trapped chernons inside SM particles.

The main results of this note are the Chern-Simons-QED$_3$ extension of the Wess-Zumino Lagrangian (\ref{chirallagrangian}), (\ref{chargedchiral}) and the viable mechanism for baryon asymmetry with the ratio $n_B/n_{\gamma} \ll 1$. Large scale cosmological simulations are needed to obtain detailed information of the properties of the model proposed above. The central experimental test of our scenario is finding no broken supersymmetry (MSSM) superpartners \cite{PDG} in the universe. 

On the theoretical side mathematical work is needed extensively. But the situation is interesting. When the Chern-Simons, or Kodama, state 
\be
\psi(A) = \mathcal{N}\exp\Big(-\frac{3}{2\LP^2\Lambda}Y_{\rm{CS}}\Big)
\label{kodama}
\ee
where $\LP$ is the Planck length and $\Lambda$ the cosmological constant and
\be
Y_{\rm{CS}} = \int A^I dA^I + \frac{1}{3}\epsilon_{IJK}A^I A^J A^K
\label{hhwf}
\ee
is reduced to mini-superspace it becomes, with some reservations, the Fourier dual of the Hartle-Hawking wave function of the universe \cite{Witten_1, Magu, Alex_H_M}.

Another interesting matter, though likewise troubled, is the possible connection of the Kodama state to quantum gravity \cite{Alex_H_F}.

\vskip 1.5cm
\appendix
\section{Chernon-particle correspondence}
\label{appdx}

The table \ref{tab:table1} gives the chernon content of SM matter and a proposal for dark matter.

\begin{table}[h!]
	\begin{center}
		\begin{tabular}{|l|l|} 
			\hline
			SM Matter & Chernon state \\ 
			\hline                                                                        
			$\nu_e$ & $m^0_R m^0_G m^0_B$ \\      
			$u_R$ & $m^+ m^+ m^0_R$ \\				 
			$u_G$ & $m^+ m^+ m^0_G$ \\  			
			$u_B$ & $m^+ m^+ m^0_B$ \\
			
			$e^-$ & $m^-_R m^-_G m^-_B$ \\		    
			$d_R$ & $m^- m^0_G m^0_B $ \\			
			$d_G$ & $m^- m^0_B m^0_R$ \\			
			$d_B$ & $m^- m^0_R m^0_G$ \\
			\hline
			Dark Matter & Chernon state  \\
			\hline
			boson (or BC) & axion(s), $s^0$ \\
			$e'$ & axino $n$ \\
			meson, baryon $o$ & $n\bar{n}, 3n$ \\
			nuclei (atoms with $\gamma ')$ & multi $n$ \\
			celestial bodies & any dark stuff \\	 
			black holes & any chernon \\
			\hline
		\end{tabular}
		\caption{\footnotesize{Visible and Dark Matter with corresponding particles. $m^0$ is color triplet, $m^{\pm}$ are color singlets. $e'$ and $\gamma '$ refer to dark electron and dark photon, respectively. BC stands for Bose condensate. Identical chernon state antisymmetrization not shown.}}
		\label{tab:table1}
	\end{center}
\end{table}


\end{document}